\documentclass{epl}

\title{Nature of quantum recurrences in coupled higher
dimensional systems}
\author{Farhan Saif\inst{1,2}}
\institute{
  \inst{1} Department of Physics, The University of Arizona, Tucson 85721, Arizona
USA.\\
  \inst{2} Department of Electronics, Quaid-i-Azam University,
Islamabad 45320, Pakistan.}


\begin{document}

\maketitle

\begin{abstract}
We investigate recurrence phenomena in coupled two degrees of
freedom systems. It is shown that an initial 
well localized wave packet displays recurrences even in the presence of coupling 
in these systems. We discuss the interdependence of
these time scales namely, classical period and quantum revival time, and explain significance of initial conditions.
\end{abstract}

Characteristics of quantum systems, which exhibit chaos in classical
domain, have posed interesting questions for researchers. In one of
these systems, namely, hydrogen atom in microwave field, the
discovery of phenomenon of quantum dynamical
localization~\cite{kn:casa} proved as a land mark in the young field
of quantum chaos. Fishman {\it et al.} connected dynamical
localization with Anderson localization~\cite{kn:fish}, and proved
it to be a generic property of periodically driven quantum systems.
Later, the phenomenon was experimentally observed~\cite{kn:arndt}.
In this paper, we develop analytical treatment for quantum
recurrence phenomena~\cite{kn:saifjopb,kn:saifee} in systems which
may exhibit chaos in classical domain and thus establish the phenomena 
generic to these systems.

Quantum recurrences originate from the simultaneous excitation of
discrete quantum levels~\cite{RobinettPR}. The existence of
recurrences has been investigated in atomic
~\cite{kn:park,kn:alber,kn:aver,kn:alber1,kn:alber2,kn:brau,kn:leic,kn:leic1,kn:stroud}
and molecular~\cite{kn:fisc,kn:vrak,kn:greb,kn:donch} wave packet
evolution. Study of some of the periodically driven quantum
systems~\cite {kn:hogg,kn:haak,kn:bres}, and two-degree-of-freedom
systems such as stadium billiard~\cite{kn:toms} indicates the
presence of quantum recurrences in higher dimensional systems as
well. Recently it is proved that the quantum recurrences are generic
to periodically driven system which may display chaos in their
classical counterpart~\cite{SaifPR}. In this paper, we study the
phenomena of dynamical recurrences in general higher dimensional
system, and provide mathematical foundations to the phenomena by
calculating recurrence times, using semiclassical secular
perturbation theory. At these time scales, namely classical period
and quantum recurrence time, an initial excitation produced in the
system displays its full or partial reappearance .

A quantum wave packet in its early evolution follows classical
mechanics and it reappears after a classical period following
classical trajectory. Later, following wave mechanics it spreads and
collapses, however, the discreteness of the quantum world leads to
its restructuring. We show that, ({\it i}) the phenomena of
dynamical recurrences occur in higher dimensional quantum systems
provided that at least two of its degrees of freedom are coupled.
({\it ii}) Furthermore we explain that, (a) the nonlinearity of the
uncoupled systems, and, (b) the initial conditions on the excitation
contribute to the classical and the quantum recurrence times
occuring in the coupled multi-dimensional systems. ({\it iii}) We
also study the interdependence of these time scales for different
kinds of dynamical systems.

We write the general Hamiltonian of a 
system with any of its two degrees of freedom coupled~\cite{kn:lieb,kn:lieb1}, as
\begin{equation}
H = H_0({\cal I})+ \lambda H_c({\cal I},{\cal \theta})
\end{equation}
where, $H_0$ is the Hamiltonian of the system in the absence of
coupling, expressed in the action coordinates ${\cal I}=(I_1, I_2)$.
Moreover, $H_c$ is the coupling Hamiltonian which describes coupling
between ${\cal I}$ and is periodic in angle, $\theta=(\theta_1,
\theta_2)$, for nonlinear resonances in the system. The parameter
$\lambda$ describes the strength of the coupling. We express the
coupling Hamiltonian as,
\begin{equation}
H_c=\sum_{n}H_{n}({\cal I})e^{i {n}.\theta},
\label{eq:hsum}
\end{equation}
where $n=(n_1, n_2)$. Whenever the frequencies $\Omega=\partial
{H_0}/\partial {\cal I}$ obey the relation,
$n.\Omega=n_1\Omega_1+n_2\Omega_2=0$, resonances occur in the
system.

We may find classical chaos in the two degrees of freedom system, as
a function of the coupling strength, whenever the two degrees of
freedom are coupled. We consider that the coupling exists between
$I_1$ and $I_2$. Within the region of resonance we find slow
variations in action, hence, following the method of secular
perturbation theory~\cite{kn:born}, we average over faster
frequency, and get the averaged Hamiltonian for the $N$th resonance,
as ${\bar H}={\bar H}_0(I)+\lambda V\cos(N\varphi)$. Here,
$\varphi=\theta_1-(M/N)\theta_2$, $I=I_1$ is the action
corresponding to the angle $\theta_1$, $V$ is the Fourier amplitude,
and $M$ and $N$ are relatively prime
integers~\cite{kn:lieb,kn:lieb1}. The action $I_2$ becomes the
constant of motion. Moreover, ${\bar H}_0(I)$ expresses uncoupled
averaged Hamiltonian. In case the coupling exists between $\cal N$
degrees of freedom, we may apply the semi-classical secular
perturbation technique $({\cal N} -1)$ times to study the effect of
most dominant degree of freedom.

The energy of the excitation changes slowly when we produce it
in the vicinity of $N$th resonance of the higher dimensional system and it is narrowly peaked.
Therefore, we expand the unperturbed energy,
${\bar H}_0(I)$, by means of Taylor's expansion around mean action $I_0$, and keep
only the terms up to second order~\cite{kn:flat}. As a
result, we express the Hamiltonian of the system, governing the evolution of the
excitation in the vicinity of $N$th resonance, as
\begin{eqnarray}
{\bar H}\cong {\bar H}_0(I_0) +\omega(I-I_0) + \frac{\zeta}{2}(I-I_0)^2+\lambda
V\cos(N\varphi).  \label{eq:reshsc}
\end{eqnarray}
Here, ${\bar H}_0(I_0)$ is the energy of the uncoupled system at the
action $I=I_0$, and $\omega$ expresses the first order derivative of
${\bar H}_0$ with respect to action calculated at $I_0$ and defines
classical frequency of the excitation. The parameter $\zeta$ defines
the nonlinear dependence of the energy of the system on the quantum
number.

We introduce the transformation $N\varphi=2\theta$ and quantize the dynamics
around the $N$th resonance by quantizing the action~\cite
{kn:berry,almeida}, that is,
\begin{equation}
I-I_0=\frac{\hbar}{i}\frac{\partial}{\partial\varphi} =\frac{N\hbar}{2i}%
\frac{\partial}{\partial\theta}.
\end{equation}
As a result, the Hamiltonian of the resonant system in the quantum domain reads
\begin{eqnarray}
{\bar H}=-\frac{N^2\zeta{\hbar}^2}{8}\frac{\partial^2} {%
\partial\theta^2}+\frac{\hbar}{2i}N\omega \frac{\partial}{\partial\theta%
} + {\bar H}_0(I_0) +\lambda V\cos 2\theta\;.  \label{eq:schham}
\end{eqnarray}
Hence, the quantum mechanical system evolves according to the
Schr\"odinger equation, ${\bar H}\psi_k={\cal E}_k\psi_k$. Here,
$\psi_k$ is the wavefunction of the system in the region of a
resonance and is, therefore, required to fulfill periodicity
condition, {\it i.e.} $\psi_k(\theta+\pi)=\psi(\theta)$. The
corresponding eigenvalue ${\cal E}_k$ defines the eigen energy of
the system. With the help of transformation,
$\psi_k=\phi_k\exp\left(-2i\omega\theta/(N\zeta \hbar)\right)$, we
map the Schr\"odinger equation on the Mathieu equation,
\begin{equation}
\left[\frac{\partial^2}{\partial\theta^2} +a_{\nu} -2q\cos 2\theta\right]\phi_{\nu}=0\;,
\label{eq:math}
\end{equation}
where,
\begin{equation}
a_{\nu}=\frac{8}{N^2\zeta{\hbar}^2}\left[{\cal E}_{\nu}-{\bar H}_0 + \frac{%
\omega^2}{2\zeta} \right]\;,  \label{8}
\end{equation}
and
\begin{equation}
q=\frac{4\lambda V}{N^2\zeta{\hbar}^2}\;.
\end{equation}
The $\pi$-periodic solutions of Eq.~(\ref{eq:math})
correspond to even functions of the Mathieu equation whose corresponding
eigenvalues are real~\cite{kn:abra}. These solutions are defined by Floquet states,
{\it i.e.} $%
\phi_{\nu}(\theta)=e^{i\nu\theta}P_{\nu}(\theta)$,
where $P_{\nu}(\theta)$ is the even order Mathieu function.

In order to obtain a $\pi$-periodic solution in $\varphi$-coordinate we
require the coefficient of $\varphi$ in the exponential factor to be equivalent
to an even integer number, $k$. This condition provides
the value for the index $\nu$ as
\begin{eqnarray}
\nu=\frac{2k}{N}+\frac{2\omega}{N\zeta\hbar}.
\label{eq:denu}
\end{eqnarray}
Therefore, we may express the eigen-energy of the system as
\begin{equation}
{\cal E}_{k}=\frac{N^2\zeta{\hbar}^2}{8}a_{\nu(k)}(q) -\frac{%
\omega^2}{2\zeta} +{\bar H}_0(I_0)\;,  \label{eq:qen}
\end{equation}
where, $a_{\nu}(q)$ is the Mathieu characteristic parameter~\cite{kn:abra}.

In order to check this result we study the case of zero coupling
strength, that is $\lambda=0$. In this case, the value for Mathieu
characteristic parameter becomes $a_{\nu}(q=0)=\nu^2$. This reduces the
quasi-energy, ${\cal E}_k$, to Eq.~(\ref{eq:reshsc}%
) in the absence of coupling term, that is, $\lambda=0$, and in addition
leads to define $k$ as $(I-I_0)/\hbar$.

The initial excitation produced at $I=I_0$ observes various time scales at which it
reappears completely or partially during its evolution.
In order to find these time scales
at which an excitation in the quantum mechanical coupled higher dimensional
system recurs, we employ the eigenenergy ${\cal E}_k$ of the system~\cite{kn:saifmg,kn:saif5}.

These time scales, $T_{\lambda}^{(i)}$,
are inversely proportional to the frequencies $\omega^{(i)}$,
where $\omega^{(i)}=(i!)^{-1}\hbar^{(i-1)} \partial^{(i)}{\cal E}_k/\partial I^{(i)}$,
when calculated at $I=I_0$. The index $i$ describes the order of
differentiation of the quasi energy ${\cal E}_k$. With the increasing values for
$i$ we have smaller frequencies indicating longer higher-order recurrence times.

The time scale, $T_{\lambda}^{(1)}=T_{\lambda}^{(cl)}$, and defines classical
period of the higher dimensional coupled system and is inversely proportional to
$\omega^{(1)}$. In the absence of coupling
$\omega^{(1)}$ reduces to $\omega$.
The time scale, $T_{\lambda}^{(2)}=T_{\lambda}^{(Q)}$,
and defines quantum mechanical recurrence time in the higher
dimensional coupled systems. It has inverse proportionality with $\omega^{(2)}$.
Here, we have
$\omega^{(2)}=(2!)^{-1}\hbar\partial^2 {\cal E}_k/\partial I^2|_{I=I_0}$.

On substituting the value for ${\cal E}_k$ from Eq.~(\ref{eq:qen}) in the defination for
$\omega^{(1)}$ and $\omega^{(2)}$,
we obtain the classical period as,
\begin{equation}
T_{\lambda}^{(cl)}=[1-M_o^{(cl)}]T_0^{(cl)},\label{eq:t1}
\end{equation}
and the quantum recurrence time for the coupled system as,
\begin{equation}
T_{\lambda}^{(Q)}=[1-M_o^{(Q)}]T_0^{(Q)}.
\label{eq:t2}
\end{equation}
Here, the time scales $T^{(cl)}_0(\equiv 2\pi\omega^{-1})$ defines
classical period and $T^{(Q)}_0 (\equiv
2\pi(\frac{1}{2!}\hbar\zeta)^{-1})$ defines quantum revival time in
the {\it absence} of coupling. The time modification factors
$M_o^{(cl)}$ and $M_o^{(Q)}$ are given as,
\begin{equation}
M_o^{(cl)}=-\frac{1}{2}\left(\frac{\lambda V \zeta} {%
\omega^2} \right)^2 \frac{1}{(1-\mu^2)^2}\, ,
\label{eq:lm1}
\end{equation}
and
\begin{equation}
M_o^{(Q)}=\frac{1}{2}\left(\frac{\lambda V \zeta} {%
\omega^2} \right)^2 \frac{3+\mu^2}{(1-\mu^2)^3}%
\,  \label{lm2}
\end{equation}
where,
\begin{eqnarray}
\mu=\frac{N\hbar\zeta}{2\omega}.
\label{mu}
\end{eqnarray}
Equations~(\ref{eq:t1}) and (\ref{eq:t2}) express
the classical period and the quantum recurrence time in the presence of coupling
between two degrees of freedom as a function of the coupling strength $%
\lambda$, nonlinearity $\zeta$ associated with the uncoupled system
and other characteristic parameters of the system. Analysis of
Eqs.~(\ref{eq:lm1}) and (\ref{lm2}) leads us to the conclusion that
the first terms of the modification factors $M_o^{(cl)}$ and
$M_o^{(Q)}$ depend quadratically on the coupling strength $\lambda$,
and, on the nonlinearity $\zeta$ present in the initial uncoupled
system. Whereas they are inversely proportional to the fourth power
of frequency $\omega$ in both the cases. The second terms are function of
$\mu$, defined in Eq.~(\ref{mu}). Hence, for the coupling
strength $\lambda=0$ we find $T_{\lambda}^{(cl)}=T_0^{(cl)}$ and
$T_{\lambda}^{(Q)}= T_0^{(Q)}$. 

{\it case a:} In the absence of nonlinearity, {\it i.e} for $\zeta=0$,
the time modification factor for the classical period
$M_o^{(cl)}$ and for the quantum recurrence time $M_o^{(Q)}$ vanish,
which is evident from Eqs.~(\ref{eq:lm1}) and (\ref{lm2}).
Thus, for linear coupled systems the quantum recurrences take
place at infinite time, {\it i.e} $T_{\lambda}^{(Q)}= T_0^{(Q)}=\infty$.
Nevertheless, the system displays recurrences after the classical period,
$T_{\lambda}^{(cl)}=T_0^{(cl)}=2\pi/\omega$. Hence, in the coupled linear
higher dimensional systems only classical periods exist.

{\it case b:} For $\mu<1$, the time modification factors $M_o^{(cl)}$ and $M_o^{(Q)}$ are related as,
\begin{equation}
M_o^{(Q)}=-3 M_o^{(cl)}=3\alpha,
\label{moq}
\end{equation}
where
\begin{equation}
\alpha= \frac{1}{2}\left(\frac{%
\lambda V \zeta}{{\omega}^2}\right)^2.
\label{alphaq}
\end{equation}

Thus, the classical period $T_{\lambda}^{(cl)}$ and the
quantum recurrence time $T_{\lambda}^{(Q)}$ in the presence of coupling are
related with the $T_0^{(cl)}$ and
$T_0^{(Q)}$ of the
uncoupled system as
\begin{equation}
\frac{3T_{\lambda}^{(cl)}}{4T_0^{(cl)}}+\frac{T_{\lambda}^{(Q)}}{4T_0^{(Q)}}=1.
\label{eq:r1}
\end{equation}
In the absence of coupling we find
$T_{\lambda}^{(cl)}=T_0^{(cl)}$ and $T_{\lambda}^{(Q)}=T_0^{(Q)}$ which
fulfills Eq.~(\ref{eq:r1}).

The case may be achieved for a weak nonlinearity in the
dominant degree of freedom of the coupled system, that is $\zeta << 1$.
The quantum recurrence time in coupled system $T_{\lambda}^{(Q)}$
and in uncoupled system $T_{0}^{(Q)}$, depend inversely on nonlinearity in the system, and are
therefore much larger than classical periods $T_{\lambda}^{(cl)}$ and $T_0^{(cl)}$ in this case.

As it follows from Eq.~(\ref{moq}), the quantum recurrence time $T_{\lambda}^{(Q)}$ reduces
by $3\alpha T_0^{(Q)}$, whereas, the classical period $T_{\lambda}^{(cl)}$ increases by
$\alpha T_0^{(cl)}$. Hence, we may conclude that the quantum dynamical recurrence
time $T_{\lambda}^{(Q)}$
reduces much faster than the classical period $T_{\lambda}^{(cl)}$ in presence of a
small nonlinearity in the system.

Since, the modification factor $\alpha$ displays direct proportionality with the
square of the nonlinearity parameter, $\zeta^2$, in the
system. In the asymptotic limit, {\it i.e.} for $\zeta$ approaching zero,
we get the result $T_{\lambda}^{(cl)}=T_0^{(cl)}$ and $T_{\lambda}^{(Q)}=T_0^{(Q)}=\infty$
as discussed in {\it case a}.

{\it case c:} In the presence of a relatively larger value of the nonlinearity parameter and/or
highly quantum mechanical systems, we may consider $\mu>1$.
We have the time modification factors related as
\begin{equation}
M_o^{(Q)}=M_o^{(cl)}=-\beta,
\label{relm}
\end{equation}
where
\begin{equation}
\beta=  \frac{1}{2}\left(\frac{4
\lambda V}{N^2 \zeta \hbar^2}\right)^2.
\label{betaq}
\end{equation}
Thus, the classical period $T_{\lambda}^{(cl)}$ and the
quantum recurrence time $T_{\lambda}^{(Q)}$ in the presence of coupling are
related with the $T_0^{(cl)}$ and $T_0^{(Q)}$ of the uncoupled system as
\begin{equation}
\frac{T_{\lambda}^{(cl)}}{T_0^{(cl)}}-\frac{T_{\lambda}^{(Q)}}{T_0^{(Q)}}=0.
\label{eq:r2}
\end{equation}

The time modification factors for the classical period and
the quantum recurrence time vanish as $\beta$ reduces to zero.
Therefore, Eqs.~(\ref{eq:t1}) and (\ref{eq:t2}) provide us the asymptotic result,
that is, $T_{\lambda}^{(cl)}=T_{0}^{(cl)}$ and
$T_{\lambda}^{(Q)}=T_{0}^{(Q)}$ and we find that Eq.~(\ref{eq:r2}) holds.

The parameter $\beta$, which determines the
modification both in classical period and in quantum recurrence time, is inversely dependent on
the fourth power of the Planck's constant $\hbar$. Hence,
for highly quantum mechanical cases we note that
$\beta$ approaches zero and the times of recurrence remain unchanged.

The classical period is inversely proportional to the classical
frequency, $\omega^{(1)}$, which is controlled by the initial excitation energy.
We note that: {\it (i)}
In presence of no nonlinearity ({\it case a}),
only classical period exists
as discussed above. Hence as frequency approaches zero, the classical
period becomes infinity which is the case of an open system;
{\it (ii)} For small value of the nonlinearity parameter in {\it case b},
as frequency becomes very small
the classical period $T_{\lambda}^{(cl)}$ and quantum recurrence time
$T_{\lambda}^{(Q)}$~\cite{kn:saifen}
changes following a square law dependence on coupling strength, $\lambda$,
as we find from Eqs.~(\ref{moq}) and (\ref{alphaq}).
However, for larger values of the frequency the
times $T_{\lambda}^{(cl)}$ and $T_{\lambda}^{(Q)}$
remain unchanged;
{\it (iii)} For relatively large value of the nonlinearity parameter, as in {\it case c},
the situation is opposite.
For smaller value of frequency we find larger value for $N$, therefore,
the modification factor reduces to zero. Hence, from Eqs.~(\ref{relm})
and (\ref{betaq}), we find that
$T_{\lambda}^{(cl)}= T_0^{(cl)}$ and $T_{\lambda}^{(Q)}= T_0^{(Q)}$.
However, for larger values of the frequency,
we note that $T_{\lambda}^{(cl)}$ and $T_{\lambda}^{(Q)}$ vary following
square law behavior as a function of modulation strength, $\lambda$.

\begin{figure}
\caption{The change in recurrence
phenomena for the wave packet originating from the center of a
resonance: (a) We plot the square of auto-correlation function,
$C^2$, as a function of time, $t$, for a Gaussian wave packet. The
wave packet is initially propagated from the center of a resonance
in atomic Fermi Accelerator~\cite{kn:saifen} and its mean position
and mean momentum are $z_0=14.5$ and $p_0=1.45$, respectively.  We
find that the wave packet experiences recurrence after each
classical period. Thick line corresponds to our numerically obtained
result and dashed line indicates quantum recurrence occuring for an
oscillator in harmonic motion. The classical period is calculated to
be $4\pi$ when the coupling strength, $\lambda=0.3$. As the coupling
strength $\lambda$ becomes zero the evolution of the wave packet
completely changes. (b) We plot square of the auto-correlation
function for the wave packet in the absence of coupling, {\it i.e.}
$\lambda=0$. We find that now the wave packet experiences collapse
after many classical periods. Later, it displays recurrence at
quantum recurrence time $T_0^{(Q)}$. The inset displays the short
time evolution of the wave packet comprising many classical periods
for $\lambda=0$. } \label{fg:rev12}
\end{figure}
For an excitation produced at the center of
a resonance the quantum evolution of the system is characterized by
Eq.~(\ref{eq:schham}) under the consideration of $\theta\approx 0$. Hence,
the secular Hamiltonian of the coupled system for exact resonance case is
\begin{eqnarray}
{\bar H}\approx -\frac{N^2\zeta{\hbar}^2}{8}\frac{\partial^2}{%
\partial\theta^2} -2 \lambda V\theta^2\;.  \label{eq:schham1}
\end{eqnarray}
Equation~({\ref{eq:schham1}}) describes the Hamiltonian of a harmonic oscillator.
Hence, this provides an evidence that if the excitation originates
initially from the center of a resonance it will
experience recurrences after each classical period, as shown in Fig. 1(a). The
quantum recurrence time now is $T_{\lambda}^{(Q)}= T_0^{(Q)}=\infty$. Thus the
evolution is the same in every dynamical system in the case of exact resonance.
The oscillator frequency is $N\sqrt{\zeta\lambda V}$. Thus
in case the coupling strength, $\lambda$, reduces to zero the harmonic
oscillator behavior disappears. The system now possesses only the
classical period $T_{0}^{(cl)}$ and the quantum recurrence time $T_0^{(Q)}$
of the uncoupled system, we find this behavior in Fig. 1(b).
This effect provides us information about level
spacing around the center of a resonance~\cite{kn:saife} as well.
For harmonic oscillator the spacing between successive levels is equal,
hence we conclude that for quasienergy levels of the Floquet operator
belonging to the center of resonance the level spacing is always equal.

We conclude that in higher dimensional systems a coupling results in modifying 
the recurrence times available for the uncoupled systems so far as the dynamics is
considered close to a nonlinear resonance. The phenomena is helpful to improve the 
efficiency of the Recurrence Tracking Microscope~\cite{rtm}. The current results may help to
identify~\cite{kn:saifee} quantum acceleration modes (QAM)~\cite{arcy} and understand the dynamics associated. Moreover 
the suggested treatment may reveal the understanding of the dynamics of Bose Einstein 
Condensates in coupled systems~\cite{martin}.


\begin{thebibliography}{999}
\bibitem{kn:saife}  Detailed numerical investigation is present in reference~\cite{kn:saifee}.
\end{thebibliography}
\end{document}